\documentclass[structabstract]{aa}  
\usepackage{graphicx}
\usepackage{natbib}
\usepackage{txfonts}
\newcommand{\lae}{\mathrel{<\kern-1.0em\lower0.9ex\hbox{$\sim$}}}
\begin{document}
   \title{The hard X-ray emission of Cen A\thanks{Based on observations with {\it INTEGRAL}, an ESA project with instruments and science data centre funded by ESA member states (especially the PI countries: Denmark, France, Germany, Italy, Switzerland and Spain), the Czech Republic, and Poland and with participation of Russia and the US.}}

%   \subtitle{I. Overviewing the $\kappa$-mechanism}

   \author{V. Beckmann
          \inst{1}
          \and
          P. Jean\inst{2,3}
	  \and
	  P. Lubi\'nski\inst{4}
	  \and
	  S. Soldi\inst{5}
	  \and
          R. Terrier\inst{1}
           %\fnmsep\thanks{Just to show the usage
           % of the elements in the author field}
          }

   \institute{Fran\c{c}ois Arago Centre, APC, Universit\'e Paris Diderot, CNRS/IN2P3, CEA/DSM, Observatoire de Paris, 13 rue Watt, 75205 Paris Cedex 13, France\\
              \email{beckmann@apc.univ-paris7.fr}
	      %\thanks{}
         \and
%         Centre d'\'Etude Spatiale des Rayonnements (CESR), OMP, UPS, CNRS;
%   B.P. 44346, 31028 Toulouse Cedex 4, France
          Universit\'e de Toulouse, UPS-OMP, IRAP, Toulouse, France
	 \and 
          CNRS, IRAP, 9 Avenue du Colonel Roche, B.P. 44346, 31028 Toulouse Cedex 4, France
	 \and
         Centrum Astronomiczne im. M. Kopernika, Rabia\'nska 8, PL-87-100 Toru\'n, Poland
	 \and
         Laboratoire AIM - CNRS - CEA/DSM - Universit\'e Paris Diderot (UMR 7158), CEA Saclay, DSM/IRFU/SAp, 91191 Gif-sur-Yvette, France\\
             %\email{c.ptolemy@hipparch.uheaven.space}
             %\thanks{The university of heaven temporarily does not
             %         accept e-mails}
             }

   \date{Received October 29, 2010; accepted April 21, 2011}

% \abstract{}{}{}{}{} 
% 5 {} token are mandatory
 
  \abstract
  % context heading (optional) leave it empty if necessary  
   {The radio galaxy Cen~A has been detected all the way up to the TeV energy range. This raises the question about the dominant emission mechanisms in the high-energy domain.}
% shown to be a simple or broken power-law like continuum and escaped to show a significant expenonential cut-off throughout the years. Nevertheless, the spectrum in the gamma-rays indicated for years that a turnover in the spectrum is required. Here, for the first time, we are able to pin-point this turnover precisely} 
  {Spectral analysis allows us to put constraints on the possible emission processes. Here we study the hard X-ray emission, in order to distinguish between a thermal and a non-thermal inverse Compton process.}
   {Using hard X-ray data provided by INTEGRAL, we determined the cut-off of the power-law spectrum in the hard X-ray domain (3 -- 1000 keV). In addition, INTEGRAL data are used to study the spectral variability. The extended emission detected in the gamma-rays by {\it Fermi}/LAT is investigated using the data of the spectrometre SPI in the 40 -- 1000 keV range.}
%   {All available INTEGRAL data are analysed. The spectrometre SPI on-board INTEGRAL provides the means to study the extended emission of Cen A in the 40 -- 1000 keV range.}
   {The hard X-ray spectrum of Cen A shows a significant cut-off at energies $E_C = 434 {+106 \atop -73}$ keV with an underlying power-law of photon index $\Gamma = 1.73 \pm 0.02$. A more physical model of thermal Comptonisation (compPS) gives a plasma temperature of $kT_{e} =$ 206$\pm 62 \rm \, keV$ within the optically thin corona with Compton parameter $y = 0.42 {+0.09 \atop -0.06}$. The reflection component is significant at the $1.9\sigma$ level with $R = 0.12 {+0.09 \atop -0.10}$, and a reflection strength $R > 0.3$ can be excluded on a $3\sigma$ level. Time resolved spectral studies show that the flux, absorption, and spectral slope varied in the range $f_{3-30 \rm \, keV} = 1.2 - 9.2 \times 10^{-10} \rm \, erg \, cm^{-2} \, s^{-1}$, $N_{\rm H} = 7 - 16 \times 10^{22} \rm \, cm^{-2}$, and $\Gamma = 1.75 - 1.87$. Extending the cut-off power-law or the Comptonisation model to the gamma-ray range shows that they cannot account for the high-energy emission. On the other hand, a broken or curved power-law model can also represent the data, therefore a non-thermal origin of the X-ray to GeV emission cannot be ruled out.
The analysis of the SPI data provides no sign of significant emission from the radio lobes and gives a $3\sigma$ upper limit of $f_{40 - 1000 \rm \, keV} \la 1.1 \times 10^{-3} \rm \, ph \, cm^{-2} \, s^{-1}$. }
%The analysis of the SPI data provides no sign of significant emission from the radio lobes in the south ($3 \sigma$ upper limit $f_{40 - 1000 \rm \, keV} \la 1.1 \times 10^{-3} \rm \, ph \, cm^{-2} \, s^{-1}$) and in the north with $f = (1.0 \pm 0.4) \times 10^{-3} \rm \, ph \, cm^{-2} \, s^{-1}$ for a non-significant signal on the $2.5 \sigma$ level. }
   {While gamma-rays, as detected by CGRO and Fermi, are caused by non-thermal (jet) processes, the main process in the hard X-ray emission of Cen~A is still not unambiguously determined, since it is either dominated by thermal inverse Compton emission or by non-thermal emission from the base of the jet.}


   \maketitle
%
%________________________________________________________________

\section{Introduction}
   
   The radio galaxy Centaurus A (NGC 5128) is probably the best-studied active galactic nucleus (AGN). With a redshift of $z = 0.001825$ and with a distance of $d = 3.8 \pm 0.1 \rm \, Mpc$ \citep{Harris10} the object is the brightest AGN at hard X-rays with $f_{20 - 100 \rm \, keV} \simeq 6 \times 10^{10} \rm \, erg \, cm^{-2} \, s^{-1}$ \citep{BeckmannINTAGN2}. For a review on Cen~A, see \cite{Israel98}. Owing to its brightness, \object{Cen A} has been a target for all main X-ray missions throughout the years, starting from the first $3 \sigma$ detection during a balloon flight in 1969 \citep{Bowyer70}. Early hard X-ray observations by {\it Ginga} and balloon-borne detectors indicated a power-law slope of $\Gamma \sim 1.8$ and a possible break at $\sim 180 \rm \, keV$ \citep{Miyazaki96}, and a break was also detected in the spectrum taken by {\it HEAO-1} at $140 \rm \, keV$ \citep{Baity81}. The {\it Compton Gamma-Ray Observatory's} ({\it CGRO}; \citealt{CGRO}) OSSE and COMPTEL instruments derived a more complex structure with a two-fold broken power-law, with breaks at $E_1 = 150 {+30 \atop -20} \rm \, keV$ and at $E_2 = 17 {+28 \atop -16} \rm \, MeV$ \citep{Steinle98}. In contrast, {\it CGRO}/BATSE observations failed to find a break in the spectrum \citep{Wheaton96,Ling00}. Also more recent {\it RXTE} and {\it INTEGRAL} observations show an absorbed unbroken power-law with photon index $\Gamma \simeq 1.8$ and an intrinsic hydrogen column density of $N_{\rm H} \simeq 10^{23} \rm \, cm^{-2}$ \citep{Rothschild06,Rothschild11}. 
% Cen~A was the only non-blazar AGN significantly detected by {\it CGRO}/COMPTEL \citep{Steinle98}.
Nevertheless, Cen~A shows a steeper spectral slope in the MeV range ($\Gamma_\gamma = 2.3 \pm 0.1$, \citealt{Collmar99}) than in the hard X-rays ($\Gamma_X \ll 2$), thus a turnover has to occur somewhere between 100 keV and a few MeV \citep{Steinle10}. The source was also the only non-blazar AGN detected by {\it CGRO}/EGRET \citep{Hartman99}, again showing a steepening of the spectrum with $\Gamma_{0.03 - 10 \rm GeV} = 2.4 \pm 0.3$ \citep{Sreekumar99}, consistent with the recent observations by {\it Fermi}/LAT ($\Gamma_\gamma = 2.7 \pm 0.1$, \citealt{Fermi_CenA_core2010}) and extending up to the TeV range as seen by the {\it HESS} experiment ($\Gamma_{E>100 \rm \, GeV} = 2.7 \pm 0.5$, \citealt{Aharonian09}).
{\it Fermi}/LAT observations also discovered extented gamma-ray emission that coincides with the radio lobes of this FR-I galaxy \citep{Fermi_CenA_extended2010}. 

% This emission is about XX times weaker than that of the central region. 
The multi-year database of the {\it INTEGRAL} mission \citep{INTEGRAL} allows us now to search the hard X-ray spectrum of Cen~A for the expected turnover. In addition, we can test whether the emission model applicable for the X-ray domain can also explain the emission at gamma-rays. 
We derived an upper limit on the hard X-ray emission from the radio lobes with {\it INTEGRAL}/SPI data and verified whether these values are consistent with the predictions based on the {\it Fermi}/LAT data.

%__________________________________________________________________

\section{Data analysis}

In this study we use all {\it INTEGRAL} data on Cen A taken between March 3, 2003 and February 21, 2009, i.e. all observations dedicated to this radio galaxy.
We used data of the imager IBIS/ISGRI \citep{Lebrun03} in the 20--1000 keV band, IBIS/PICsIT \citep{Labanti03} data at $234 - 632 \rm \, keV$, and those of the spectrometre SPI \citep{Vedrenne03} between 40 keV and 1850 keV. The two JEM-X monitors \citep{Lund03} provide spectral information in the 3 -- 30 keV band.
We selected the data up to an off-axis angle of $10^\circ$ for IBIS/ISGRI and SPI, and within $3^\circ$ for JEM-X. Data reduction was performed using the Offline Scientific Analysis (OSA) package version 9.0 provided by the ISDC \citep{Courvoisier03}. The PICsIT data analysis followed the procedure described in \cite{Lubinski09}. The different instruments were not always switched on simultaneously. This is mainly because the two JEM-X monitors are operated one at a time in order to spare the instrument and for telemetry constraints. In addition, the SPI spectrometre undergoes periods of annealing to repair the crystal structure of the GeD detectors, so was not available during all observations. The effective exposure times are therefore different, with 145 ks for JEM-X1, 96 ks JEM-X2, 1,425 ks IBIS/ISGRI, $2,076 \rm \, ks$ IBIS/PICsIT, and 1,858 ks for SPI. 

Errors quoted in this work are at the 90\% confidence level, corresponding to $\sim 1.6 \sigma$.

\section{The hard X-ray spectrum}

The fitting and modelling  of the X-ray spectra were performed using XSPEC version 12.5.1 \citep{Arnaud96}. Only the data of JEM-X are significantly influenced by the intrinsic absorption, whereas for IBIS/ISGRI and SPI the absorption value was fixed to the average value of $N_{\rm H} = 10^{23} \rm \, cm^{-2}$ (e.g. \citealt{Grandi03}, \citealt{Benlloch01}).
An estimate of the turnover in the spectrum of Cen~A can already be achieved by analysing the IBIS/ISGRI data alone. A simple absorbed power-law model with $\Gamma = 1.84$ gives a reduced chi-square of $\chi^2_\nu = 4.8$ for 31 degrees of freedom. Adding a high-energy cutoff improves the fit to $\chi^2_\nu = 1.1$ (30 d.o.f.). The best-fitting values are a photon index of $\Gamma = 1.67 \pm 0.03$ and cut-off energy of $E_C = 298 {+56 \atop -42} \rm \, keV$. 
Another phenomenological model that includes a steepening of the spectrum is the broken power-law. The fit result is only slightly worse than the cut-off power-law with $\chi^2_\nu = 1.1$ (28 d.o.f.). The low-energy photon index of this model is $\Gamma_1 = 1.74 \pm 0.02$ with a break energy of $E_{\rm break} = 48 {+6 \atop -5} \rm \, keV$ and a high-energy slope of $\Gamma_2 = 1.95 {+0.03 \atop -0.03}$.

\begin{figure}
   \centering
   \includegraphics[angle=90,width=8.3cm]{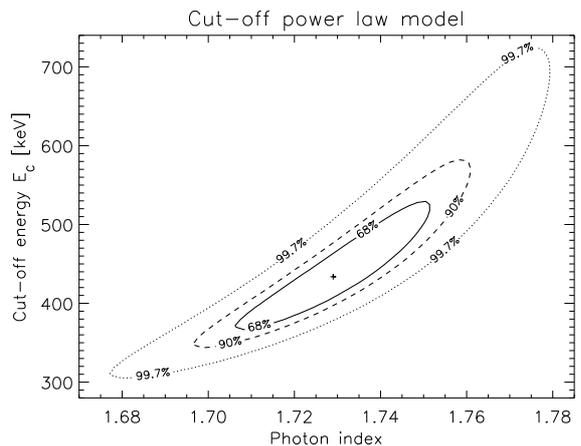}
   \caption{68\% ($1\sigma$), 90\% ($1.6 \sigma$), and 99.7\% ($3\sigma$) confidence level contours of the high-energy cut-off energy $E_C$ versus the photon index $\Gamma$ based on combined {\it INTEGRAL} data. The best fit value is indicated by a "+".}
              \label{fig:cutoff_contour}%
    \end{figure}

Cen~A is known to maintain its spectral shape under flux variations (e.g. \citealt{Rothschild06}, and this work Section~\ref{timeresolved}). We therefore added all {\it INTEGRAL} spectra together, i.e. JEM-X1, JEM-X2, SPI, IBIS/PICsIT, and IBIS/ISGRI.
Also here, a simple absorbed power-law ($\Gamma = 1.85 \pm 0.01$) does not provide a good representation of the data ($\chi^2_\nu = 1.83$, 104 d.o.f.). The absorbed cut-off power-law model gives $\chi^2_\nu = 1.06$ (103 d.o.f.) with a photon index of $\Gamma = 1.73 \pm 0.02$ and $E_C = 434 {+106 \atop -73} \rm \, keV$. The spectral shape and the cut-off energy are not independent variables. This can be seen in the contour plot of $E_C$ and $\Gamma$ as shown in Fig.~\ref{fig:cutoff_contour}. A flatter spectrum is compensated for in the fit by a lower cut-off energy. Nevertheless, a cut-off energy below 300 keV and above 700 keV can be ruled out at a 99.7\% confidence level. The spectral slope is better constrained by the data spanning 3 -- 1000 keV. Even when considering different cut-off energies, the $3\sigma$ range for the photon index is $1.68 < \Gamma < 1.78$. The broken power-law fit gives an equally good representation of the data with $\chi^2_\nu = 1.06$ (102 d.o.f.). The turnover appears at an energy of $E_{\rm break} = 50 {+13 \atop -6} \rm \, keV$, with a spectral slope of $\Gamma_1 = 1.78 {+0.02 \atop -0.02}$ and $\Gamma_2 = 1.95 {+0.05 \atop -0.03}$ below and above the break, respectively. 

The cut-off power-law and the broken power-law model are phenomenological descriptions of the spectrum. A physical model should include the processes of thermal Comptonisation. Here we apply the {\tt compPS} model \citep{Poutanen96}. We assume that the plasma of temperature $T_e$ forms an infinite slab on top of the accretion disk, with the plasma having an optical depth $\tau$. Since $T_e$ and $\tau$ are intrinsically strongly anticorrelated, we fit the Compton parameter $y = 4 \tau kT_e / (m_e \, c^2)$ (with $m_e$ % being 
the electron mass) instead of the optical depth.  The temperature of the seed photons is not well constrained by the hard X-ray spectrum used here, thus we assume a multicolour disk with a fixed inner temperature $T_{bb} = 10 \rm \, eV$. The fit to the {\it INTEGRAL} data results in a reduced chi-squared of $\chi^2_\nu = 1.02$ for 101 d.o.f.

\begin{figure}
   \centering
   \includegraphics[angle=270,width=9cm]{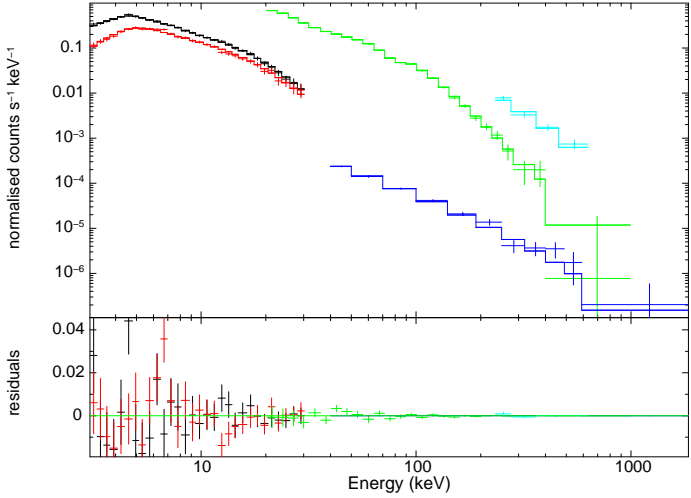}
   \caption{Count spectrum of Cen~A using {\it INTEGRAL} JEM-X1 and JEM-X2 ($3-30 \rm \, keV$), IBIS/ISGRI (20 -- 1000 keV), IBIS/PICsIT (234 -- 632 keV) and SPI data ($40-1400 \rm \, keV$). The residuals in the lower panel are with respect to the best-fitting Comptonisation model ({\tt compPS}).}
    \label{fig:countspec}
    \end{figure}

Figure~\ref{fig:countspec} shows the count spectrum of Cen~A along with the fit residuals. It is worth noting that the iron K$\alpha$ fluorescence line at $E_{\rm K\alpha} = 6.39 \rm \, keV$ is only visible in the JEM-X2 spectrum. Analysing the JEM-X spectra individually confirms this observation. The JEM-X1 spectrum gives a $3\sigma$ upper limit of $f_{\rm K\alpha} < 2 \times 10^{-3} \rm \, ph \, cm^{-2} \, s^{-1}$, corresponding to an equivalent width of $EW < 330 \rm \, eV$. For the JEM-X2 data set, the same analysis gives $f_{\rm K\alpha} = (9 \pm 5) \times 10^{-4} \rm \, ph \, cm^{-2} \, s^{-1}$, corresponding to an equivalent width of $EW = (250 \pm 110) \rm \, eV$. This is consistent with measurements taken by {\it RXTE} over 12.5 years, which show equivalent widths in the range $50 - 160 \rm \, eV$ \citep{Rothschild11}.

\begin{figure}
   \centering
   \includegraphics[angle=90,width=9cm]{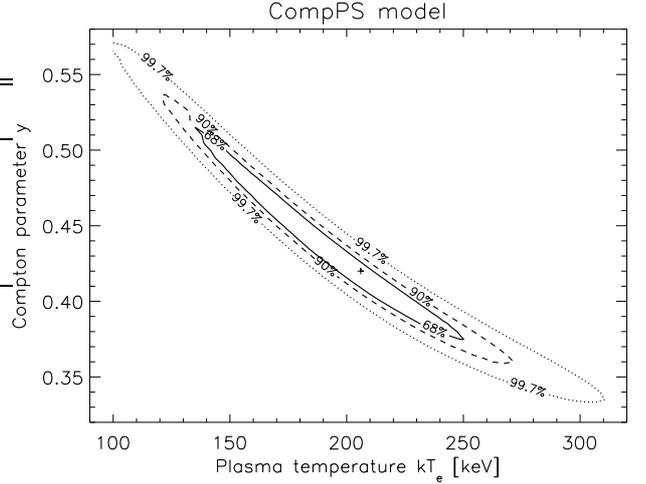}
   \caption{Contour plot for the {\tt compPS} model fit of the {\it INTEGRAL} data of Cen~A. Both, Compton parameter $y$ and plasma temperature $kT_e$ are constrained.}
              \label{fig:compps_contour}
    \end{figure}

\begin{table}
\caption{Spectral fits to combined {\it INTEGRAL} data. All errors are at the $90\%$ confidence level.}             % title of Table
\label{fitsummary}      % is used to refer this table in the text
\centering                          % used for centering table
\begin{tabular}{c c c c c c}        % centered columns (4 columns)
\hline\hline                 % inserts double horizontal lines
model & $\Gamma$ & $E_C$ or $kT_e$ & y & R & $\chi^2_\nu$ (d.o.f.)\\ 
      &          &      [keV]      &   &   &        \\   % table heading 
\hline                        % inserts single horizontal line
   power & $1.85\pm 0.01$  & -- & -- & -- & 1.83 (104)\\
   cut-off & $1.73 \pm 0.02$ & $434 {+106 \atop -73}$ & -- & -- & 1.06 (103)\\
   bkn po & $1.78 / 1.95$ & $50 {+13 \atop -6}$ & -- & -- & 1.06 (102)\\
   pexrav & $1.75 \pm 0.04$ & $549 {+387 \atop -168}$ & -- & $0.07 {+0.11 \atop -0.07}$ &  1.07 (102)\\
   compPS & -- & $206 \pm 62$ & $0.42 {+0.09 \atop -0.06}$ & $0.12 {+0.08 \atop -0.10}$ & 1.02 (101) \\
\hline                                   %inserts single line
\end{tabular}
\end{table}
The {\tt compPS} model gives an electron plasma temperature of $kT_e = 206 \pm 62 \rm \, keV$, and a Compton parameter $y = 0.42 {+0.09 \atop -0.06}$ corresponding to an optical depth of $\tau = 0.26$. In this fit, we had to let the plasma temperature of the PICsIT spectrum to be independent from the other instruments, as the PICsIT spectrum appears slightly harder. For the PICsIT spectrum this results in a temperature as high as $675 \rm \, keV$, which cannot be constrained by the fit.
Similar to the tight connection between the cut-off energy and the underlying spectral slope in the case of the cut-off power-law model, in the {\tt compPS} model the plasma temperature $T_e$ and the Compton parameter $y$ are linked to each other. Figure~\ref{fig:compps_contour} shows the contour plot for $y$ versus $kT_e$ with 68\%, 90\%, and 99.7\% confidence levels. Considering all combinations, the plasma temperature range is very wide ($300 \rm \, keV > kT_e > 100 \rm \, keV$) with a Compton parameter $0.34 < y < 0.57$ corresponding to a range of optical depth $0.14 < \tau < 0.69$. 

\begin{figure}
   \centering
   \includegraphics[angle=90,width=8cm]{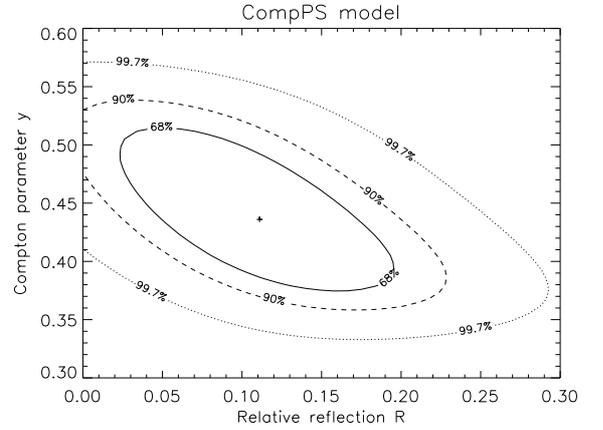}
   \caption{Contour plot for the {\tt compPS} model fit of the {\it INTEGRAL} data of Cen~A. The reflection component is only significant on the $1.9 \sigma$ level, with a $3\sigma$ upper limit of $R \lae 0.3$.}
              \label{fig:R-y-contour}
    \end{figure}

The reflection strength is $R = 0.12 {+0.08 \atop -0.10}$, inconsistent with there being no reflection at the $1.9\sigma$ level. Also the reflection strength $R$ is linked to the Compton parameter $y$. The contour plot in Fig.~\ref{fig:R-y-contour} shows the connection between $R$ and $y$. While the possibility that no significant reflection is detectable in Cen~A cannot be excluded, the $3\sigma$ upper limit of $R \lae 0.3$ considering all combinations of $R$ and $y$ is a strong constraint. In the course of the studies of NGC~4151 \citep{Lubinski10} we investigated the influence of cross-calibration issues on the strength of the detected reflection component. Comparison between reflection values derived from combined {\it XMM-Newton} and {\it RXTE} data versus {\it INTEGRAL} spectra showed no significant difference. In the case of Cen~A the spectrum is also not as strongly impacted by absorption as NGC~4151. Problems in earlier versions of the OSA software, where JEM-X spectra would drop to energies above 10 keV, are also not apparent in the spectrum as shown in Fig.~\ref{fig:countspec}. Thus, we do not expect cross-calibration to cause an artificially enhanced reflection component.
The average model flux in the X-rays as detectable by {\it INTEGRAL} is $f_{(3 - 1000 \rm \, keV)} = 2.7 \times 10^{-9} \rm \, erg \, cm^{-2} \, s^{-1}$, and thus Cen~A displays a luminosity of $L_X = 2.0 \times 10^{43} \rm \, erg \, s^{-1}$. 

For comparison we applied the simpler {\tt pexrav} model, which describes an exponentially cut-off power-law spectrum reflected from neutral material, with the output spectrum being the sum of the cut-off power-law and the reflection component \citep{Magdziarz95}. % Also applying this model the plasma temperature appears higher if we use the cut-off energy of the spectrum as an indicator. 
In this case the best-fit results ($\chi^2_\nu = 1.07$ for 102 d.o.f.) are a photon index of $\Gamma = 1.75 \pm 0.04$, cut-off energy $E_C = 549 {+387 \atop -168} \rm \, keV$, and relative reflection of $R = 0.07 {+0.11 \atop -0.07}$, consistent with no reflection component. Thus, the {\tt pexrav} model does not require a reflection component at all, with a $3\sigma$  upper limit of $R \lae 0.28$, close to the value derived from the {\tt compPS} model fit. It should be taken into account that the results of these two models are expected to differ slightly. The model {\tt compPS} takes into account Comptonisation of the reflected component, whereas {\tt pexrav} assumes that there is no reprocessing in the plasma region.

All the fit results concerning the combined {\it INTEGRAL} spectrum are summarised in Table~\ref{fitsummary}. The number of fitted parameters $n$ for each model is $n=8$ (absorbed power-law model,  {\tt power}), $n = 9$ (absorbed cut-off power-law, {\tt cut-off}), $n=10$ (absorbed broken power-law, {\tt bkn po}), $n = 10$ (absorbed Comptonisation model, {\tt pexrav}), and $n = 11$ (absorbed Comptonisation model, {\tt compPS}). 

\section{Time resolved spectroscopy}
\label{timeresolved}

As pointed out before, the {\it INTEGRAL} data analysed in the overall spectrum are not simultaneous. Therefore, uncertainty about the normalisation might influence the results. 
To test whether the spectrum indeed maintains its shape under flux variations, we analysed combined JEM-X and ISGRI data that were truly simultaneous. We grouped data together into 12 simultaneous JEM-X and ISGRI spectra with approximately 20 ks each. We applied a simple absorbed power-law model to the combined data, using the same normalisation for JEM-X and ISGRI. The photon indices were in the range $1.75 < \Gamma < 1.87$ with the average value of $\langle \Gamma \rangle = 1.81$. Nine of the values are consistent with the average value on the $1\sigma$ level, the remaining three photon indices on the $2 \sigma$ level.  When fixing the photon index to the average value during the fit, all fits still give reasonable fit results ($\chi^2_\nu < 1.3$, with an average $\langle \chi^2_\nu \rangle = 1.0$). The intrinsic absorption shows significant variations in the range $7.5 \times 10^{22} \rm \, cm^{-2} < N_{\rm H} < 16.4 \times 10^{22} \rm \, cm^{-2}$ with an average value of $\langle N_{\rm H} \rangle = 11 \times 10^{22} \rm \, cm^{-2}$. It has to be taken into account, though, that these variations do not influence the spectrum above a few keV and thus should not change the results on the reflection component and/or cut-off of the hard X-ray spectrum. No significant correlation is found between the X-ray flux and the hydrogen column density.  

We then also tested the spectral and flux variability using the single science window JEM-X spectra, which last about half an hour each. The results were similar to those in the combined JEM-X and ISGRI spectral timing analysis but with a larger scatter. The $3-20 \rm \, keV$ photon indices were in the range $1.5 < \Gamma < 2.4$ with the average value of $\langle \Gamma \rangle = 1.85 \pm 0.18$ when fixing the intrinsic absorption to $N_{\rm H} = 11 \times 10^{22} \rm \, cm^{-2}$. For a fixed photon index of $\Gamma = 1.81$, the absorption was in the range $3 \times 10^{22} \rm \, cm^{-2} < N_{\rm H} < 24 \times 10^{22} \rm \, cm^{-2}$ with an average of $\langle N_{\rm H} \rangle = (11 \pm 4) \times 10^{22} \rm \, cm^{-2}$. If both parameters were left free to vary, the result was $1.4 < \Gamma < 2.6$ and $3 \times 10^{22} \rm \, cm^{-2} < N_{\rm H} < 37 \times 10^{22} \rm \, cm^{-2}$. Also here, no correlation can be detected between flux and intrinsic absorption.

\section{Extended emission}

As for a radio galaxy, Cen~A (RA $= 201.3651^\circ$, DEC= $-43.0191^\circ$) is known to have radio lobes extending out from the core by about 300~kpc. These lobes are located in the north at RA$ = 201.5^\circ$, DEC $= -39.8^\circ$ and in the south at RA = $200.625^\circ$, DEC = $-44.716^\circ$. Recently, {\it Fermi}/LAT detected emission from positions consistent with these lobes at significance levels of $5\sigma$ (1FGL~J1333.4--4036) and $8\sigma$ (1FGL~J1322.0--4545) for the northern and southern lobes, respectively \citep{Fermi_CenA_extended2010}. Here we used the data of the spectrometre SPI to test for emission of the lobes at hardest X-rays. Because of its wide field of view and a resolution of $2.5^\circ$, SPI is able to map large-scale structures on the hard X-ray background. 
% we used: Cen A: 201.3651, -43.01911, Cen A north: 201.5, -39.8, Cen A south: 200.625, -44.716
% 11 other sources in the LAT field, e.g. 1FGL J1307.0-4030, 13 07 06, -40 30 37
The spectra of the radio lobes and Cen A were extracted by model fitting, assuming that the sky intensity distribution consist of these three point sources, at the positions mentioned above.  We performed the analysis in 10 keV wide energy bins, covering the 40--1850 keV range. Each energy bin is adjusted to the data for each germanium detector separately, assuming that the count rate is due to the sum of the sky contributions and the instrumental background. The latter is assumed to be proportional to the rate of saturating events in germanium detectors \citep{Jean03}. The spectra were rebinned in logarithmic spaced energy bins for the spectral analysis presented in Section~3.

The analysis of the SPI data shows no sign of significant emission from the radio lobes in the south ($3 \sigma$ upper limit $f_{40 - 1000 \rm \, keV} \la 1.1 \times 10^{-3} \rm \, ph \, cm^{-2} \, s^{-1}$) and in the north with $f = 1.0 \pm 0.4 \times 10^{-3} \rm \, ph \, cm^{-2} \, s^{-1}$ for a non-significant signal on the $2.5 \sigma$ level. 
   \begin{figure}
   \centering
   \includegraphics[width=9cm]{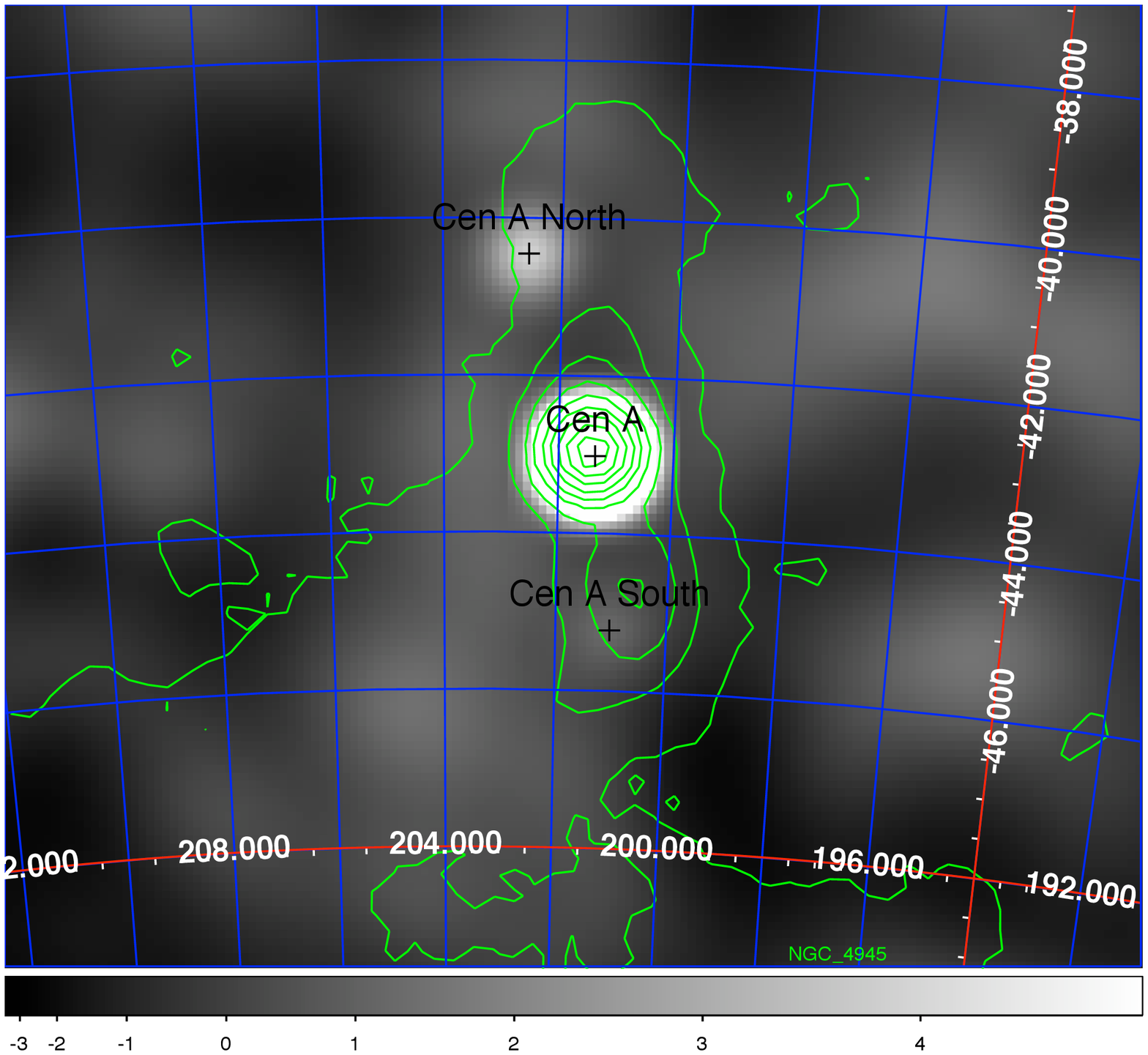}
   \caption{{\it INTEGRAL}/SPI significance map in the 40 -- 1000 keV energy band of the Cen~A region. Overlaid are the contours of the WMAP radio map. %Note that the hard X-ray emission of the lobes is not significant.
   } 
              \label{fig:SPI_map}
    \end{figure}
We investigated the excess in the north of Cen~A further by fitting a source model in various energy bands to the northern lobe in which position, extension, and orientation angle of a two-dimensional assymmetric Gaussian were left free to vary. In this case we find a excess of $f_{80 - 400 \rm \, keV} = (9.1 \pm 7.8) \times 10^{-4} \rm \, ph \, cm^{-2} \, s^{-1}$ ($3\sigma$ error) at RA = $202.9^\circ$, DEC = $-40.2^\circ$. The $3.5\sigma$ signal is too low to claim a significant detection at this point. Even if we consider this flux to be a reliable measurement, the hard X-ray lobe emission of Cen~A can still be lower, as other sources in the field can also contribute, e.g. in the north 1FGL~J1307.0--4030 detected by {\it Fermi}/LAT, or another so far undetected blazar in the background. Based on the SPI data, it is not yet possible to determine whether the excess we detect is extended or resulting from a point-like source.

If we consider the flux of $f = 10^{-3} \rm \, ph \, cm^{-2} \, s^{-1}$ an upper limit, this value agrees with the expected energy distribution presented by \cite{Fermi_CenA_extended2010}. At an energy of 300 keV, they expect a contribution to the SED of $\sim 5 \times 10^{12} \rm \, Jy \, Hz$ in the case of the southern lobe, whereas the upper limit based on the SPI data corresponds to $\la 10^{13}  \rm \, Jy \, Hz$. For the northern lobe, the emission expected from the Fermi/LAT extrapolation is even lower. Therefore, unless about five times more SPI data will be accumulated on Cen~A, the lobes will most likely escape detection at hardest X-rays. 

\section{Discussion}

In the first approximation, the average Cen~A spectrum in the 3 -- 1000 keV band as measured by {\it INTEGRAL} can be represented by an absorbed simple power-law model with $\Gamma = 1.85$ and intrinsic absorption in the range $6 \times 10^{22} \rm \, cm^{-2} < N_{\rm H} < 12 \times 10^{22} \rm \, cm^{-2}$, although the fit has a bad quality. The spectral slope is consistent with earlier measurements by {\it BeppoSAX} and {\it RXTE}, giving absorption values of $N_{\rm H} \sim 10^{23} \rm \, cm^{-2}$ and $\Gamma = 1.80$ above 3 keV \citep{Grandi03,Benlloch01}. The slope is slightly flatter than reported from early {\it INTEGRAL} measurements with $\Gamma = 2.02 \pm 0.03$ \citep{Soldi05}, $\Gamma = 1.94 \pm 0.02$ \citep{BeckmannINTAGN1}, and $\Gamma = 1.94 \pm 0.07$ \citep{Rothschild06}. \cite{Rothschild06} point out that the spectra get flatter with the on-going improvement of the calibration and software of the {\it INTEGRAL} mission, i.e. that the discrepancy might have been non-physical. The situation seemed to have been resolved with software versions 7 and higher of OSA: the spectral slope of Cen~A in the second {\it INTEGRAL} AGN catalogue \citep{BeckmannINTAGN2} using OSA~7 is the same as presented here.

A good fit to the data is achieved by an absorbed cut-off power-law model. Here, for the first time using {\it INTEGRAL} data, we have detected a significant cut-off at energies $E_C = 434 {+106 \atop -73} \rm \, keV$, with photon index $\Gamma = 1.73 \pm 0.02$. These values are consistent with previous attempts to determine the cut-off using {\it RXTE} data, which resulted in lower limits of $E_C$ in the range 400 keV to 1600 keV \citep{Rothschild06}. A recent study of {\it RXTE} data again finds no sign of a cut-off, but puts the lower limit at $E_C > 2 \rm \, MeV$ \citep{Rothschild11}. Not detecting a cut-off led to the conclusion that the X-ray range in Cen~A might be jet-dominated, rather than originating from thermal inverse Compton processes. Support for this scenario was given by the fact that no significant reflection component could be found.
In the work presented here we indeed see such a reflection component when applying a Compton reflection model ({\tt compPS}), although the reflection strength is comparably low ($R = 0.12 {+0.08 \atop -0.10}$), and inconsistent with no reflection only on the $1.6\sigma$ level. The $3 \sigma$ upper limit of $R \lae 0.3$ indicates that the reflection component, if present in Cen A, has to be weak. 
The plasma temperature is $kT_E = 206 \pm 62 \rm \, keV$ with a Compton parameter $y = 0.42 {+0.09 \atop -0.06}$. It has to be pointed out, though, that these two parameters are closely linked and that the resulting values also depend on the choice of the seed photon temperature. If we assume here an inner disk temperature of $T_{bb} = 100 \rm \, eV$ instead of $T_{bb} = 10 \rm \, eV$, the plasma temperature rises to $kT_e = 320 {+40 \atop -62} \rm \, keV$ and the Compton parameter reduces to $y = 0.24 {+0.06 \atop -0.03}$. On the other hand, measurements of other sources derived $T_{bb}$ values between 10 eV and 30 eV \citep{Petrucci01,Lubinski10}.

\begin{figure}
   \centering
   \includegraphics[height=9cm,angle=270]{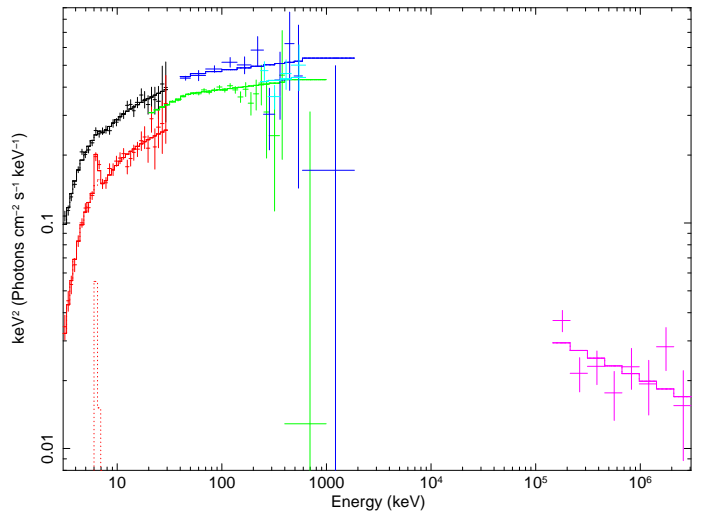}
   \caption{Combined {\it INTEGRAL} and (non-simultaneous) {\it Fermi}/LAT spectrum of Cen~A in $\log E \times f_E$ versus $\log E$. The data are modelled by a double broken power-law with individual normalisation for the data of different epochs. The highest energy bins for IBIS/ISGRI and SPI are only upper limits.} 
              \label{fig:CenA_Fermi}
    \end{figure}

The perfect match of the Comptonisation model to the {\it INTEGRAL} data seems to rule out the dominant emission in the X-rays being non-thermal, i.e. also arising from the jet as the gamma-ray emission. We have shown, though, that also a simple broken power-law and a cut-off power-law model represent the data.
After extrapolation of the cut-off power-law model, the {\tt compPS} or the {\tt pexrav} model into the {\it Fermi}/LAT energy range shows that an inverse Compton component dominating the X-rays cannot be responsible for the gamma-ray emission. All these models predict no detectable flux at energies above 100 MeV. On the other hand, the broken power-law model connects to the GeV emission rather smoothly. In Fig.~\ref{fig:CenA_Fermi} we show the {\it INTEGRAL} spectrum, together with the spectrum derived from {\it Fermi}/LAT data. 
%   \begin{figure}
%   \centering
%   \includegraphics[height=9cm,angle=270]{bkn2po_Fermi_ufspec.ps}
%   \caption{Combined {\it INTEGRAL} and {\it Fermi}/LAT spectrum of Cen~A. The data are modeled by a double broken power-law with individual normalisation for the data of different epochs.} 
%              \label{fig:CenA_Fermi}
%    \end{figure}
The LAT spectrum was extracted from {\it Fermi} data from the early mission in August 2008 until January 2011, applying standard extraction methods and considering the flux contribution by the extended lobes. We derive a photon index of $\Gamma_\gamma = 2.8 \pm 0.1$ for the energy range 150~MeV to 3~GeV and flux $f_{>100 \rm MeV} = 2.3 \times 10^{-7} \rm \, ph \, cm^{-2} \, s^{-1}$, consistent with earlier results from the {\it Fermi} collaboration ($\Gamma_\gamma = 2.7 \pm 0.1$, $f_{>100 \rm MeV} = 2.2 \times 10^{-7} \rm \, ph \, cm^{-2} \, s^{-1}$ \citealt{Fermi_CenA_core2010}. The high-energy spectrum of Cen~A seems to have another break in the range 500 keV to 100 MeV. In fact, the combined fit of {\it INTEGRAL} and LAT data with an absorbed broken power-law gives a break energy at $E_{\rm break} = 103 \rm \, keV$ and photon indices of $\Gamma_1 = 1.8$ and $\Gamma_2 = 2.4$, below and above the break, respectively, with a reduced $\chi^2$ value of $1.25$ (109 d.o.f.). A double broken power-law fit gives a significantly better result. Here, the breaks occur at $E_{b1} = 49 {+9 \atop -4} \rm \, keV$ and $E_{b2} = 97 {+144 \atop -65} \rm \, MeV$, with the three power-law slopes of $\Gamma_1 = 1.78 {+0.01 \atop -0.02}$,  $\Gamma_2 = 1.95 \pm 0.03$, and  $\Gamma_3 = 2.4 {+0.2 \atop -0.1}$, for a $\chi^2_\nu = 1.04$ (107 d.o.f.). An F-test shows that the improvement is highly significant compared to the broken power-law model. % 5e-5
As the energy range between 500 keV and 100 MeV is not covered, the second break is not strongly constrained. Under the assumption of non-thermal inverse Compton emission, one would expect a continuous curvature, which the broken power-law model only represents roughly. 

In the following we discuss the two options. First we consider the hard X-ray emission to be thermal Comptonisation and thus typical of a Seyfert type galaxy. Then, we investigate the implications when the hard X-rays are dominated by non-thermal emission.

\subsection{Cen A as a Seyfert galaxy}

If we assume that thermal Comptonisation is the dominant mechanism in the hard X-rays, an additional component, e.g. non-thermal (jet) emission would have to be present in order to explain the gamma-rays. This is well-established (e.g. \citealt{Johnson97}, \citealt{Fermi_CenA_core2010}). The presence of thermal Comptonisation is also supported by the iron line and the Compton reflection hump, if we consider the $1.9 \sigma$ detection. \cite{Fermi_CenA_core2010} have presented already a model considering the combined thermal inverse Compton along with a non-thermal gamma-ray component to explain the X-ray to VHE emission.

The results can be compared to those obtained for other bright Seyfert galaxies with high signal-to-noise, hard X-ray data. \object{NGC~4151} is an object of comparable brightness, distance category ($z = 0.0033$), and spectral type (Seyfert 1.5). Applying the same {\tt compPS} model with $T_{bb} = 10 \rm \, eV$, \cite{Lubinski10} derive plasma temperatures in the range $50 \rm \, keV < kT_e < 200 \rm \, keV$ with Compton parameters $1.2 > y > 0.6$ and corresponding optical depth $0.6 < \tau < 0.3$ while the source is undergoing different flux states. In this context, the values of Cen~A we derive here are similar to the dim state in NGC~4151 ($kT_e \simeq 200 \rm \, keV$, $y \simeq 0.9$, $\tau \simeq 0.4$), although this state has a much stronger reflection component ($R \simeq 0.8$). Taking the 3 -- 1000 keV X-ray luminosity as a proxy for the bolometric luminosity, under the assumption that the synchrotron branch emits about as much as the inverse Compton branch, gives a bolometric luminosity of $L_{bol} \simeq 4 \times 10^{43} \rm \, erg \, s^{-1}$. By extracting the 2--10 keV luminosity ($L_{2-10 \rm keV} = 2.7 \times 10^{42} \rm \, erg \, s^{-1}$) and applying the conversion factor for the bolometric luminosity $L_{bol}/L_{2-10 \rm keV} = 9$ presented by \cite{Marconi04}, we derive a value of $L_{bol} = 2.4 \times 10^{43} \rm \, erg \, s^{-1}$. It has to be taken into account that this is assumes that the spectrum in the hard X-rays is based on thermal Compton emission. Considering that there is also a non-thermal component, the value of the bolometric luminosity we give here has to be understood as an upper limit for the thermal accretion luminosity.
With a central black hole mass of $M_{BH} \simeq 5 \times 10^7 \rm \, M_\odot$ \citep{Neumayer07}, this translates into an Eddington ratio around $\lambda = 0.01$, similar to NGC~4151 with $\lambda \simeq 0.01 - 0.02$ \citep{Lubinski10}. Based on long-term X-ray measurements by {\it RXTE}, \cite{Rothschild11} derived a lower average luminosity of $L_{bol} = 8 \times 10^{42} \rm \, erg \, s^{-1}$ and thus an Eddington ratio of only $\lambda \simeq 0.001$. That Cen~A is detectable in gamma-rays, while NGC~4151 is not, is the result of the strong jet-component apparent in the former. It also shows that the presence of a jet in Cen~A has no strong influence on the central engine, which appears to have similar X-ray characteristics to those in NGC~4151. 

\cite{Petrucci01} used {\it BeppoSAX} data of 6 bright Seyfert galaxies to determine their hard X-ray properties, finding plasma temperatures in the range $170 \rm \, keV < kT_e < 315 \rm \, keV$ with optical depths $0.2 > \tau > 0.05$ and an average reflection strength of $R \simeq 1$. Again, in comparison Cen~A appears to represent a rather dim state of a Seyfert, which is accompanied by a higher plasma temperature.

The measured cut-off energy of $E_C = 434 {+106 \atop -73} \rm \, keV$ is, however, consistent with the average cut-off observed in Seyfert~2 galaxies by {\it BeppoSAX}. \cite{Dadina08} find for this AGN type an average of $E_C = 376 \pm 42 \rm \, keV$ with a photon index of $\Gamma = 1.80 \pm 0.05$. The reflection strength of $R = 0.1$ is rather low compared to the average of $R=1$ as found by {\it BeppoSAX} \citep{Dadina08} and {\it INTEGRAL} \citep{BeckmannINTAGN2} for large samples of Seyfert galaxies.

A different picture has been drawn from {\it CGRO}/OSSE results. Here, the average cut-off energy of Seyfert~2 appeared at $E_C = 130 {+220 \atop -50}$ keV with a spectral slope of $\Gamma = 1.33 {+0.56 \atop -0.52}$ \citep{Zdziarski00}. The analysis with the Comptonisation model {\tt compPS} gave an average temperature of only $kT_e = 84 {+101 \atop -31} \rm \, keV$ and a Compton parameter as high as $y = 1.1 {+0.3 \atop -0.4}$ resulting in an optical depth of $\tau = 1.7$. This effect of an increasing plasma temperature with decreasing Compton parameter can be seen in Fig.~\ref{fig:compps_contour}, but the OSSE results are significantly different from those obtained by {\it BeppoSAX} and {\it INTEGRAL}. One reason might be the higher energy band for OSSE, starting at 50 keV, whereas {\it INTEGRAL} (3 -- 1000 keV) and {\it BeppoSAX} (1 -- 150 keV) probe a softer energy range. Thus, OSSE data do not constrain the photon index of the spectrum before the high-energy cut-off as strongly as the instruments of {\it BeppoSAX} and {\it INTEGRAL} do.

When we combine all these results and assume a thermal inverse Compton component, the hard X-ray emission of Cen~A could indicate that we observe here a typical Seyfert~2 AGN core, although slightly at the dim end of the distribution, which is connected to a higher plasma temperature, and thus a higher spectral cut-off than observed on average. 
This in turn would indicate that the non-thermal component apparent in the gamma-rays and detected by {\it CGRO}/EGRET and {\it Fermi}/LAT has no effect on or direct connection to the inverse Compton component: the hard X-ray spectrum of Cen~A does not appear to be significantly different from Seyfert galaxies, which do not show strong jet emission.

%{\tt \cite{Roustazadeh11}: modelling of Cen A with gamma-ray induced pair cascades}

\subsection{The Cen~A core as a FR~I with inefficient accretion flow}

Cen~A shows properties of both a Seyfert~2 galaxy and an FR-I galaxy. In the central regions one observes a jet at a large angle to the line of sight. The jet appears to be relativistic up to a distance of $\sim 1.5 \rm \, pc$ from the core. In addition, the radio core indicates that all elements for a BL Lac type core are observed in Cen~A. Thus, the X-ray band could also be dominated by non-thermal emission of a misaligned blazar. There are several observational facts that favour this interpretation. 
For an absorbed Seyfert core with $N_{\rm H} \simeq 10^{23} \rm \, cm^{-1}$, the lack of a significant reflection component ($R \ll 0.3$) and the weak iron line with $EW \simeq 250 \rm \, eV$ are certainly unusual. It has to be pointed out, though, that also other bright and well-studied Seyfert galaxies, such as \object{NGC~4593} and \object{NGC~4507}, show spectra with no reflection component, and that for several others the reflection strength is detectable only at a $\lae 2 \sigma$ level (Lubi\'nski et al., in prep.). If we assume a BL Lac type central engine in Cen~A, the X-rays up to the gamma rays would be produced by inverse Compton upscattering in the jet, possibly of synchrotron photons produced in the jet itself via synchrotron self Compton (SSC) processes. 
% In this case, the iron line could be produced in the circumnuclear disk at $r \sim 150 \rm \, pc$ and no reflection component would be expected. 
As the AGN core has moderate luminosity, accretion in Cen~A appears to be radiatively inefficient with an Eddington ratio around $\lambda \simeq 0.01$. \cite{Hardcastle07} argue that in these cases almost all the available energy from the accretion process could be channelled into the jets. The dominant accretion mechanism could be Bondi accretion \citep{Bondi52}, which is direct infall from the hot, X-ray emitting medium without a disk forming around the core. This would naturally explain the lack of the reflection component in Cen~A. 

\cite{Marchesini04} find a dichotomy in radio galaxies with respect to the accretion efficiency. Objects with % $L_{bol} \lae 10^{44} \rm \, erg \, s^{-1}$ and 
inefficient accretion ($\lambda \lae 0.1$) like Cen~A, would be dominated by non-thermal emission, while the more efficient accreting FR-II galaxies would have a higher ratio in accretion versus jet power. \cite{Balmaverde06} have studied the X-ray properties of FR-I galaxies. They show that, when assuming a non-thermal origin for the X-ray cores in FR-I at the base of the jet, only about $\lambda \sim 10^{-7}$ would be emission related directly to accretion and not be channelled into the jet. Thus, in an FR-I like Cen~A the non-thermal component should clearly dominate the Seyfert-type emission at high energies. \cite{Allen06} come to a similar conclusion by studying X-ray luminous elliptical galaxies, in which Bondi type accretion can account for the observed emission and where most of the accreted matter would enter the jets, giving a close correlation between $L_{Bondi}$ and $L_{jet}$. 

The hard X-ray to gamma-ray spectrum, such as shown in \cite{Steinle98} and in this work, is indeed a curved continuum, with a photon index $\Gamma \simeq 1.8$ in the X-rays, $\Gamma \simeq 2.3$ in the MeV range, and $\Gamma \simeq 2.8$ at energies $E > 100 \rm \, MeV$. As described in \cite{Fermi_CenA_core2010}, the entire SED can be modelled by a one-zone SSC model, although the authors show that the combination of a thermal inverse Compton component with a non-thermal jet emission fits the SED equally well. 

% NGC 5548, BeppoSAX (Dadina 2008), OSSE (Zdziarski, Poutanen \& Johnson 2000)
% Cen A jet has complex X-ray structure: while the inner part of the jet is dominated by knots and has properties consistent with local particle acceleration at shocks, the particle acceleration in the outer 3.4 kpc of the jet is likely to be dominated by an unknown distributed acceleration mechanism \cite{Hardcastle07b}. They also confirm that the jet knots do not show dramatic flux variations, different from e.g. M87.

A difficulty of the non-thermal model is to explain the very persistent and stable emission from the Cen~A core. X-ray flux and spectrum vary only slightly, as shown in this work, and stronger variability, both in luminosity and in spectral slope, would be expected from a beamed source. It also has to be pointed out that an SSC model for Cen~A can represent the X-ray to GeV spectrum, but fails to explain the TeV emission detected by HESS \citep{Fermi_CenA_core2010}, so that in this case another model component would also be necessary.

Another possibility is that in Cen~A both components, the thermal and the non-thermal inverse Compton component, are observable in the X-ray band. A similar case supposedly exists in 3C~273. According to studies of {\it Beppo}/SAX spectra of this blazar by \cite{Grandi04}, the jet component is highly variable, while the underlying Seyfert core component is persistent. This leads to a variable ratio of non-thermal versus thermal emission from the core of 3C~273. 

\section{Conclusions}

The extended emission detected by {\it Fermi}/LAT in gamma rays  cannot be seen in {\it INTEGRAL}/SPI data, with a $3\sigma$ upper limit of the emission of $f_{40 - 1000 \rm \, keV} < 0.001 \rm \, ph \, cm^{-2} \, s^{-1}$ for the northern and southern lobes. The upper limit is consistent with the expected energy distribution following the analysis of \cite{Fermi_CenA_extended2010}.  

Using the combined {\it INTEGRAL} data the presence of curvature in the hard X-ray to gamma domain is now evident, which can be described by an exponential cut-off or break in the spectrum. It is possible to determine the cut-off energy $E_C \simeq 400 \rm \, keV$
%and plasma temperature 
in Cen~A, confirming earlier claims about a turnover based on {\it CGRO} data. A non-thermal origin to the hard X-ray emission cannot be ruled out though.
A thermal Comptonisation model gives a good representation of the data ($\chi^2_\nu = 1.02$). Applying a Comptonisation model, we find an electron plasma with temperature $kT_e = 206 \rm \, keV$ and optical depth $\tau = 0.26$ ($y = 0.42$). The Comptonisation model, and the simpler cut-off power-law, cannot explain the emission measured at gamma rays, which has to be due to an additional component, e.g. the jet. 
On the other hand, interpreting the hard X-ray spectrum as non-thermal (e.g. SSC) emission from the jet would explain the low reflection component we measure ($R \ll 0.3$) and the relatively weak iron line. In addition, the smooth transition from the X-rays to the MeV and GeV region can indicate that the non-thermal component dominates the whole high-energy range, yet the very stable X-ray spectrum is a challenge for a model based on jet-emission.

From {\it CGRO}/COMPTEL observations we know that the transition to the gamma-ray emission is indeed smooth. This means that a thermal Compton component, which would decline rapidly towards higher energies, would have to be compensated for by an increasingly important non-thermal component already at energies of a few MeV, because it is the dominant emission process at $E > 1 \rm \, MeV$. A similar situation could be present in other {\it Fermi}/LAT detected sources that are not bona-fide blazars; i.e., the hard X-rays are dominated by thermal Comptonisation, but the non-thermal component sets in the low MeV domain, peaking at some tens of MeV. A Compton telescope like {\it CAPSiTT} \citep{CAPSiTT}, covering this energy range (1 MeV $< E < 100 \rm \, MeV$), would be crucial for verifying either of the two scenarios.

\begin{acknowledgements}
       We thank the anonymous referee for the comments that helped to improve the paper. This research has made use of the NASA/IPAC Extragalactic Database (NED), which is operated by the Jet Propulsion Laboratory, California Institute of Technology, under contract with the National Aeronautics and Space Administration. We thank Peter Kretschmar for providing the script for the contour plots. PL has been supported by the Polish MNiSW grants NN203065933 and 362/1/N-INTEGRAL (2009-2012).
\end{acknowledgements}

\bibliographystyle{aa}
\bibliography{CenA.bib}

\end{document}